\documentclass[preprint,preprintnumbers,amssymb,amsmath,9pt, superscriptaddress]{revtex4}[10pt]
\usepackage{hyperref}
\usepackage{epsfig,graphicx,psfrag}
\usepackage{dcolumn}
\usepackage{bm}
\usepackage{comment}
\begin{document}
\input epsf.tex
\newcommand{\beq}{\begin{eqnarray}}
\newcommand{\eeq}{\end{eqnarray}}
\newcommand{\nn}{\nonumber}
\def\ltap{\ \raise.3ex\hbox{$<$\kern-.75em\lower1ex\hbox{$\sim$}}\ }
\def\gtap{\ \raise.3ex\hbox{$>$\kern-.75em\lower1ex\hbox{$\sim$}}\ }
\def\CO{{\cal O}}
\def\CL{{\cal L}}
\def\CM{{\cal M}}
\def\tr{{\rm\ Tr}}
\def\CO{{\cal O}}
\def\CL{{\cal L}}
\def\CM{{\cal M}}
\def\mpl{M_{\rm Pl}}
\def\Z{{\bf Z}}
\newcommand{\bel}[1]{\be\label{#1}}
\newcommand{\be}{\begin{equation}}
\newcommand{\ee}{\end{equation}}
\newcommand{\bea}{\begin{eqnarray}}
\newcommand{\eea}{\end{eqnarray}}
\newcommand{\p}{\partial}
\def\al{\alpha}
\def\bt{\beta}
\def\eps{\epsilon}
\def\eg{{\it e.g.}}
\def\ie{{\it i.e.}}
\def\mn{{\mu\nu}}
\newcommand{\rep}[1]{{\bf #1}}
\def\be{\begin{equation}}
\def\ee{\end{equation}}
\def\bea{\begin{eqnarray}}
\def\eea{\end{eqnarray}}
\newcommand{\eref}[1]{(\ref{#1})}
\newcommand{\Eref}[1]{Eq.~(\ref{#1})}
\newcommand{\gsim}{ \mathop{}_{\textstyle \sim}^{\textstyle >} }
\newcommand{\lsim}{ \mathop{}_{\textstyle \sim}^{\textstyle <} }
\newcommand{\vev}[1]{ \left\langle {#1} \right\rangle }
\newcommand{\bra}[1]{ \langle {#1} | }
\newcommand{\ket}[1]{ | {#1} \rangle }
\newcommand{\ev}{{\rm eV}}
\newcommand{\kev}{{\rm keV}}
\newcommand{\Mev}{{\rm MeV}}
\newcommand{\gev}{{\rm GeV}}
\newcommand{\tev}{{\rm TeV}}
\newcommand{\mev}{{\rm MeV}}
\newcommand{\meV}{{\rm meV}}
\newcommand{\mnu}{\ensuremath{m_\nu}}
\newcommand{\nnu}{\ensuremath{n_\nu}}
\newcommand{\mlr}{\ensuremath{m_{lr}}}
\newcommand{\acc}{\ensuremath{{\cal A}}}
\newcommand{\mav}{MaVaNs}
\newcommand{\h}{{\cal H}}
\newcommand{\hb}{{\cal \bar H}}
\def\draftnote#1{{\bf #1}}
\newcommand{\bpl}{\ensuremath{B+L}\ }
\newcommand{\bml}{\ensuremath{B-L}\ }
\newcommand{\ct}{\ensuremath{c_\theta}}
\newcommand{\st}{\ensuremath{s_\theta}}

\title{The Dark Side of the Electroweak Phase Transition}
\author{Subinoy Das}
\affiliation{Department of Physics and Astronomy, University Of British Columbia,
BC, V6T 1Z1 Canada}
\author{Patrick J. Fox}
\affiliation{Theoretical Physics Department, Fermilab, Batavia, IL 60510, USA }
\author{Abhishek Kumar}
\affiliation{Center for Cosmology and Particle Physics,
  Dept. of Physics, New York University,
New York, NY 10003}
\author{Neal Weiner}
\affiliation{Center for Cosmology and Particle Physics,
  Dept. of Physics, New York University,
New York, NY 10003}
\preprint{\small FERMILAB-PUB-09-090-T}
\date{\today}
\begin{abstract}
Recent data from cosmic ray experiments may be explained by a new GeV scale of physics. In addition the fine-tuning of supersymmetric models may be alleviated by new $\mathcal{O}$(GeV) states into which the Higgs boson could decay.  The presence of these new, light states can affect early universe cosmology.  We explore the consequences of a light ($\sim$ GeV) scalar on the electroweak phase transition.  We find that trilinear interactions between the light state and the Higgs can allow a first order electroweak phase transition and a Higgs mass consistent with experimental bounds, which may allow electroweak baryogenesis to explain the cosmological baryon asymmetry.  We show, within the context of a specific supersymmetric model, how the physics responsible for the first order phase transition may also be responsible for the recent cosmic ray excesses of PAMELA, FERMI etc.
We consider the production of gravity waves from this transition and the possible detectability at LISA and BBO.  
\end{abstract}

\maketitle

\section{Introduction}
\label{sec:intro}

The central unresolved question of high energy physics is the nature of electroweak symmetry breaking. Many issues are raised by this: what stabilizes the weak scale against radiative corrections? Is there a fundamental Higgs or is it composite, or is the symmetry breaking due to strong dynamics?

The nature of the electroweak phase transition may be particularly important for our universe. If the electroweak phase transition is first order, it provides one of the essential conditions for baryogenesis \cite{Cohen:1990py,Cohen:1993nk}. The transition must be strongly first order to prevent washout of the generated baryon number by sphalerons in thermal equilibrium.  Within the standard model~\footnote{In addition, the CP violating phase in the SM, present in the CKM matrix, is too small to generate sufficient baryon number.  In this paper we concentrate only on the order of the phase transition and assume the other conditions~\cite{Sakharov:1967dj} necessary for baryogenesis are satisfied.} achieving sufficient strength requires a Higgs lighter than $\sim 42~\gev$, strongly excluded by existing LEP2 Higgs searches \cite{Carena:1996wj,Barate:2003sz}.

However, in moving beyond the standard model, there is again the possibility of a first order electroweak phase transition. For instance, in supersymmetric theories, couplings to the top squark~\cite{Carena:1996wj} can allow a larger thermal cubic term in the potential, to compensate for the larger quartic associated with the Higgs mass. In the NMSSM, the presence of an additional singlet can provide a first order phase transition~\cite{Menon:2004wv}.  Similar physics can arise within the context of the singlet Majoron model for neutrino masses \cite{Cline:2009sn}.  The presence of effective dimension six Higgs operators, generated when heavy states are integrated out, also increase the strength of the phase transition~\cite{Grojean:2004xa}.

To generate a first order phase transition, there must be new physics at or below the weak scale. Because of the hierarchy problem, new fields are naturally expected at $O(M_W)$. More recently, however, there have been possible indications of new physics at the $\sim$ GeV scale. Evidence for an excess of cosmic ray positrons from PAMELA \cite{Adriani:2008zr}, coupled with the absence of anti-protons \cite{Adriani:2008zq}, as well as possible signals at ATIC \cite{:2008zzr}, PPB-BETS \cite{Yoshida:2008zz} and WMAP \cite{Finkbeiner:2003im,Finkbeiner:2004us,Dobler:2007wv} provide a motivation for a new, light scale of physics \cite{ArkaniHamed:2008qn}. The large cross section necessary to explain the excess can be understood from a Sommerfeld enhancement from some new, light force carrier \cite{ArkaniHamed:2008qn,Pospelov:2008jd}\footnote{The Sommerfeld enhancement was first explored in the context of dark matter in \cite{Hisano:2004ds}, arising from weak interactions for multi-TeV WIMPs. See also \cite{Cirelli:2008id,MarchRussell:2008yu}.}, while the hard lepton spectrum (without anti-protons) can be realized from the new mediator \cite{Cholis:2008vb,Cholis:2008qq}\footnote{Another possibility is that the annihilations occur through a force carrier that couples to leptons only \cite{Cirelli:2008pk,Fox:2008kb}, or which is leptonic, itself \cite{Zurek:2008qg}, but there must still be a light state in order to generate the large cross section.}. Finally, the INTEGRAL 511 keV excess can be explained in the XDM scenario \cite{Finkbeiner:2007kk}, which postulates the existence of a sub-GeV force carrier. While such a scale is unnatural in the standard model, it can arise naturally, for instance, in supersymmetric theories \cite{ArkaniHamed:2008qp,Baumgart:2009tn}.

Moreover, theories with additional standard model singlets can allow a Higgs lighter than the present LEP bound of 114.4 \gev\ \cite{Dermisek:2005ar,Dermisek:2005gg,Chang:2005ht,Schuster:2005py,Graham:2006tr,Chang:2006bw,Chang:2008cw}. This occurs when neutral particles much lighter than the Higgs open up additional decay modes, which are less constrained than for a standard model Higgs. In particular, the new states must be sufficiently light that they do not decay into b-quarks, for which the bounds are stronger, motivating again masses in the few GeV range or lower.  Finally, some have suggested that the excess of multi-muon events at CDF \cite{Aaltonen:2008qz} might be explained by new, light states with masses 3.6, 7.3 and 15 GeV \cite{Giromini:2008xh}.

In this paper, we consider the effects of a light scalar on the electroweak phase transition. We consider in some senses a ``minimal" model, with simply a trilinear interaction between the Higgs boson and the singlet $s$. As a consequence, when the Higgs acquires a vev, there is a mixing between the states allowing the lighter to be constrained from Higgs-strahlung searches at LEP.  We find there are regions of parameter space consistent with these LEP limits which generate the strong first order phase transition necessary for electroweak baryogenesis. All regions have $m_s \ltap 12 \gev$. This study is distinct from and complementary to that of the parameter scan of \cite{Profumo:2007wc}, where no viable models were found with states below 12 GeV. It adds to the analysis of the Majoron model for neutrino masses of \cite{Cline:2009sn} in that we are not tied to connections to neutrino mass, and we can consider scalars lighter than 5 GeV because of a broader particle content.

Models with simply a trilinear interaction between the Higgs and the singlet occur in certain scenarios of supersymmetry breaking~\cite{Dine:1992yw,Fox:2002bu}. Furthermore, if such models are extended to include dark matter (DM) the singlet may not only generate a strong electroweak phase transition but can also lead to a low velocity enhancement in the DM-DM annihilation cross section that may  explain recent cosmic-ray anomalies.

The layout of this paper is as follows: in the next section, we explain, simply, why a singlet with a trilinear gives rise to a strong first order phase transition, and study the effects on the sphaleron energy in the case when it is weakly mixed, as well as discussing LEP experimental constraints.  In section~\ref{sec:susy} we consider a supersymmetric realization of this model.  In section~\ref{sec:bubble} we consider the nucleation of bubbles in these scenarios and find the amplitude of the GW signal is probably too small to be detected in upcoming gravity wave experiments. Finally, in section~\ref{sec:conclusions} we conclude.

\section{A first order phase transition from a new light boson}
\label{sec:setup}

\begin{figure}[t] 
   \centering
   \psfrag{sinsth}[][][1.]{$\sin\theta$}
   \psfrag{ms}[][][1.]{$\overline{m}_S$}
   \psfrag{0.0}[][][.8]{$0$}
   \psfrag{0.1}[][][.8]{$\!0.1$}
   \psfrag{0.2}[][][.8]{$\!0.2$}
   \psfrag{0.3}[][][.8]{$\!0.3$}
   \psfrag{0.4}[][][.8]{$\!0.4$}
   \psfrag{0.5}[][][.8]{$\!0.5$}
   \psfrag{0}[][][.8]{$0$}
   \psfrag{2}[][][.8]{$2$}
   \psfrag{4}[][][.8]{$4$}
   \psfrag{6}[][][.8]{$6$}
   \psfrag{8}[][][.8]{$8$}
   \psfrag{10}[][][.8]{$10$}
   \psfrag{12}[][][.8]{$12$}
   \psfrag{14}[][][.8]{$14$}
   \psfrag{gev25}[][][0.45]{$25~\gev$}
   \psfrag{gev30}[][][0.45]{$30~\gev$}
   \psfrag{gev40}[][][0.45]{$40~\gev$}
   \psfrag{gev50}[][][0.45]{$50~\gev$}
   \includegraphics[width=2.5in]{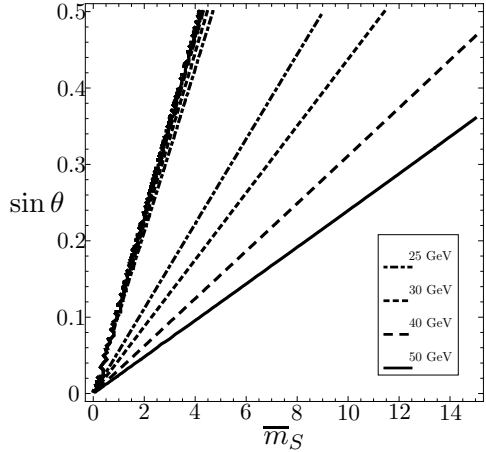}
   \caption{The critical temperature of the phase transition, for $\overline{m}_H=115~\gev$, there is little dependence on $\overline{m}_H$.  From the right the lines correspond to $T_c=$ $50~\gev$ (solid), $40~\gev$ (dashes), $30~\gev$ (short dashes), $25~\gev$ (dot-dashed) respectively.  At the boundary on the left the critical temperature grows rapidly.  Here the transition becomes a second order phase transition, the minimum near the origin is never the true minimum of the potential.}
   \label{fig:Tcplot}
\end{figure}

We consider the standard model (SM) Higgs, $\phi$, coupled to a light neutral singlet scalar, $s$.  At zero temperature the potential is
\be
\label{eqn:zerotemppot}
V=-D T_0^2\phi^2 + \frac{\lambda}{4} \phi^4
+ \frac{1}{2}m_s^2 s^2 + \kappa\, \phi^2 s~.
\ee
The parameters in the potential can be re-expressed in terms of the physical masses ($\overline{m}_H$, $\overline{m}_S$), mixing angle and vev ($v=246~\gev$) as:
\bea
D T_0^2 &= \frac{\overline{m}_H^2 \overline{m}_S^2}{4(\overline{m}_S^2 c_\theta^2 + \overline{m}_H^2 s_\theta^2)},  m_s^2 &= \overline{m}_S^2 c_\theta^2 + \overline{m}_H^2 s_\theta^2 \nonumber \\
\lambda&=\frac{\overline{m}_H^2 c_\theta^2+\overline{m}_S^2 s_\theta^2}{2v^2}, \kappa&=\frac{(\overline{m}_H^2-\overline{m}_S^2)s_{2\theta}}{4v}~.
\eea
At finite temperature, for a light Higgs, this receives corrections:\bea
V=&&D(T^2-T_0^2)\phi^2 -E T \phi^3 + \frac{\lambda_T}{4} \phi^4 \nonumber \\
&&+ \frac{1}{2}m_s^2 s^2 + \kappa\, \phi^2 s + \frac{\kappa}{12}  T^2 s~,
\eea
with $E= \frac{1}{4\pi v^3}(2m_W^3 + m_Z^3)\approx 10^{-2}$, $D=\frac{1}{8v^2}(2m_W^2+m_Z^2+2m_t^2)\approx 1/6$, $\lambda_T$ is logarithmically sensitive to $T$ \cite{Anderson:1991zb,Carena:2004ha}.  Since we will be interested in the case of small mixing ($\kappa \ll 1$) we have dropped terms higher order in $\kappa$.  This leads to a first order phase transition at a critical temperature,
\be
\label{eqn:Tcrit}
T_c=\frac{T_0}{\sqrt{1-\frac{E^2}{D\left(\lambda_T-2\frac{\kappa^2}{m_s^2}\right)} -\frac{\kappa^2}{12m_s^2 D}}}~.
\ee
The scalar vevs at the second minimum are
\be
\label{eqn:vevs}
\phi_c=\frac{2ET_c}{\lambda-2\frac{\kappa^2}{m_s^2}}\, , \quad s_c=-\frac{\kappa}{m_s^2}\left( \phi_c^2+\frac{T_c^2}{12}\right).
\ee
The presence of the light scalar alters both the critical temperature and the Higgs vev.  As we will see in the next subsection, when discussing the strength of the phase transition in this model, the figure of merit is $\phi_c/T_c$ just as it is in the SM.  The trilinear coupling increases this ratio, making the phase transition stronger.  It is also possible in this model to greatly lower the temperature of the phase transition.  In a particular region of parameter space the potential is such that the critical temperature is considerably lower than the symmetry breaking scale (Figure~\ref{fig:Tcplot}), the phase transition takes place at late times.  This is significantly lower than what occurs in the SM.  Such a late phase transition opens the possibility, if the Dark Matter (DM) is coupled to the scalar sector, of appreciably changing the properties of DM between the early universe and today~\cite{Cohen:2008nb}.

The critical temperature is shown in Figure~\ref{fig:Tcplot}, for physical Higgs mass, $\overline{m}_H=115~\gev$.  For a given singlet mass there is a maximum value of the mixing angle, beyond which there is no first order phase transition.  For mixing angles larger than this maximum the coupling between $\phi$ and $s$ is large enough that the negative $s$-vev~(\ref{eqn:vevs}), which scales with $T^2$, destabilizes the origin and the true minimum is always the symmetry breaking minimum away from the origin.  Symmetry is not restored at high temperature~\cite{Weinberg:1974hy,Mohapatra:1979vr}.

Note that the increase in the strength of the phase transition is a tree-level effect that appears in the denominator, due to the (cubic) mixing term $\kappa$ in the Lagrangian.   Since it is tree level the effect is potentially large. Alternative approaches to increasing the strength of the phase transition~\cite{Profumo:2007wc} often rely upon loop generated contributions to the terms appearing in the numerator, e.g. stop loops increasing $E$.  At the same time, this is distinct from tree-level effects of higher dimension operators \cite{Grojean:2004xa}. If one considers the decoupling limit ($m_s \gg m_h$) of this model, while keeping the first order phase transition intact with a constant $\kappa^2/m_s^2$, one finds the first order nature arises because decoupling the singlet simply suppressed the Higgs quartic directly. That is, the decoupling limit of this model, rather than yielding important higher dimension operators, merely yields a theory with a light Higgs boson, which generates a first order phase transition simply due to the light Higgs. It is only because the singlet is light that we have a consistent phenomenology of {\em both} the $O(100)$ GeV Higgs boson and the first order phase transition simultaneously.

\subsection{Effects on sphaleron energy}
\label{sec:sphaleron}

In the SM \bml is a good symmetry whereas \bpl is anomalous.  The breaking of \bpl is mediated by the sphaleron field configuration and the sphaleron energy \cite{Klinkhamer:1984di} is $E_{sph}=\frac{4\pi v(T)}{g_2}B\left(\frac{\lambda}{g^2}\right)$
where $B(x)$ is a numerical factor varying between $B(0)\approx 1.6$ and $B(\infty)\approx 2.7$.

At finite temperature this field configuration may be formed by a thermal fluctuation.  The sphaleron has size $R_{sph}\sim 1/M_W$, far larger than its Compton wavelength, and contains $\mathcal{O}(1/\alpha_W)$ fields so its probability of production is given by the classical thermal Boltzman distribution, $\Gamma\sim e^{-E_{sph}(T)/T}$.  The requirement that the net \bpl charge is not washed out leads to a constraint on the order of the phase transition, $E_{sph}/T\gtap 40$.  Using knowledge of $B\left(\frac{\lambda}{g^2}\right)$ this becomes
\be
\label{eqn:orderconstraint}
\frac{v(T)}{T}\gtap 1~.
\ee
This is the famous statement that the phase transition must be strongly first order.  Including an extra singlet coupled to the Higgs will alter the sphaleron energy and hence the constraint, which we now discuss.

We take an ansatz for the gauge fields and scalar fields of,
\bea
W_i^a\sigma^a=-\frac{2i}{ g_2}f(\xi)dU\,U^{-1}, & \phi=\frac{v}{\sqrt{2}}h(\xi) U
\begin{pmatrix}
0 \\
1
\end{pmatrix}, & s=x s(\xi)
\eea
where
\be
U=\frac{1}{r}\begin{pmatrix}
z & x+i y\cr
-x+i y & z
\end{pmatrix}
\ee
Where the asymptotic form of the profiles are
\be
\label{eq:bcs}
\begin{matrix}
f(\xi\rightarrow 0)  = f_0 \xi^2,  & h(\xi\rightarrow 0) =  h_0 \xi,  & s(\xi\rightarrow 0) = s_0 \\
\!\!\!\!\!\!\!\! f(\xi\rightarrow \infty) =  1, &  \!\!\!\! h(\xi\rightarrow \infty) = 1, &  s(\xi\rightarrow \infty) = 1
\end{matrix}
\ee

\begin{figure}[t] 
   \centering
   \psfrag{sinth}[][][1.]{$\sin\theta$}
   \psfrag{ms}[][][1.]{$\overline{m}_S$}
   \psfrag{0.0}[][][.8]{$0$}
   \psfrag{0.1}[][][.8]{$\!0.1$}
   \psfrag{0.2}[][][.8]{$\!0.2$}
   \psfrag{0.3}[][][.8]{$\!0.3$}
   \psfrag{0.4}[][][.8]{$\!0.4$}
   \psfrag{0.5}[][][.8]{$\!0.5$}
   \psfrag{0.6}[][][.8]{$\!0.6$}
   \psfrag{0.8}[][][.8]{$\!0.8$}
   \psfrag{1.0}[][][.8]{$\!1.0$}
   \psfrag{0}[][][.8]{$0$}
   \psfrag{2}[][][.8]{$2$}
   \psfrag{4}[][][.8]{$4$}
   \psfrag{6}[][][.8]{$6$}
   \psfrag{8}[][][.8]{$8$}
   \psfrag{10}[][][.8]{$10$}
   \psfrag{12}[][][.8]{$12$}
   \psfrag{14}[][][.8]{$14$}
   \includegraphics[width=2.5in]{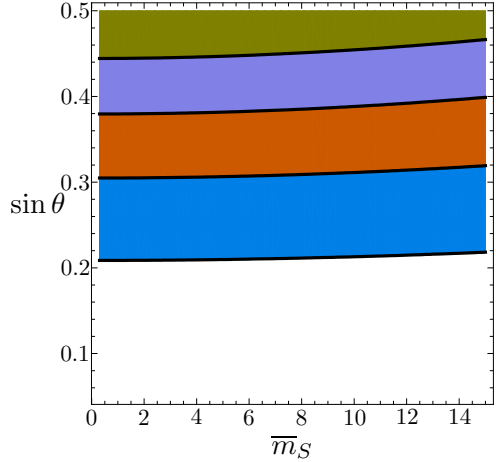}
   \caption{Sphaleron energy for $\overline{m}_H=115\,\gev$, again there is only weak dependence on Higgs mass.  From the top the contours correspond to $E_{sph}/\left(4\pi v/g\right)=1.82,\, 1.84,\, 1.86,\,1.88$ respectively. In the SM, $E_{sph}/\left(4\pi v/g\right)\approx 1.9$.}
   \label{fig:sphaleronenergy}
\end{figure}

To solve for the field profiles, and hence calculate the sphaleron energy in this model~\footnote{Note that the sphaleron solution is found in the limit where there is no mixing with $U(1)_Y$, $\theta_w =0$.  Corrections to the energy are small as one moves to the physical value \cite{Klinkhamer:1990fi,Kleihaus:1991ks}.}, we use a shooting technique.  Initial guesses are made for the profiles at small and large radii, consistent with (\ref{eq:bcs}).  These are then evolved up and down respectively and compared at some intermediate radius.  A solution is found by varying the boundary conditions and minimising the difference between the upper and lower solutions at the intermediate point.  The resulting sphaleron energy is shown in Figure~\ref{fig:sphaleronenergy}.  The addition of the singlet does not appreciably change the energy of the sphaleron configuration and so the constraint on the order of the phase transition is not greatly altered.  This is easily understood: the light scalar varies on scales larger than the size of the sphaleron so the Higgs profile acts like a delta function source to the singlet, and causes it to turn on.  Since the singlet profile is slowly varying on the length associated with the Higgs profile it may be approximated as a constant vev.  Its only effect is to shift the mass of the Higgs from its SM value, $m_\phi^2=\left(m_H^{SM}\right)^2 +2\kappa x$.    Since the sphaleron energy has a weak dependence on the Higgs mass \cite{Klinkhamer:1984di} the addition of a light singlet does not greatly alter the requirement of (\ref{eqn:orderconstraint}).

\subsection{A strongly first order phase transition and experimental constraints}
\label{sec:constraints}

\begin{figure}[t] 
   \centering
   \psfrag{sinth}[][][1.]{$\sin\theta$}
   \psfrag{ms}[][][1.]{$\overline{m}_S$}
   \psfrag{0.0}[][][.8]{$0$}
   \psfrag{0.1}[][][.8]{$\!0.1$}
   \psfrag{0.2}[][][.8]{$\!0.2$}
   \psfrag{0.3}[][][.8]{$\!0.3$}
   \psfrag{0.4}[][][.8]{$\!0.4$}
   \psfrag{0.5}[][][.8]{$\!0.5$}
   \psfrag{0}[][][.8]{$0$}
   \psfrag{2}[][][.8]{$2$}
   \psfrag{4}[][][.8]{$4$}
   \psfrag{6}[][][.8]{$6$}
   \psfrag{8}[][][.8]{$8$}
   \psfrag{10}[][][.8]{$10$}
   \psfrag{12}[][][.8]{$12$}
   \psfrag{14}[][][.8]{$14$}
   \psfrag{alpgtptfive}[][][.8]{$\alpha>0.5$}
   \psfrag{mheq90}[][][.8]{$\overline{m}_H=90\,\gev$}
   \psfrag{mheq115}[][][.8]{$\overline{m}_H=115\,\gev$}
   \includegraphics[width=2.5in]{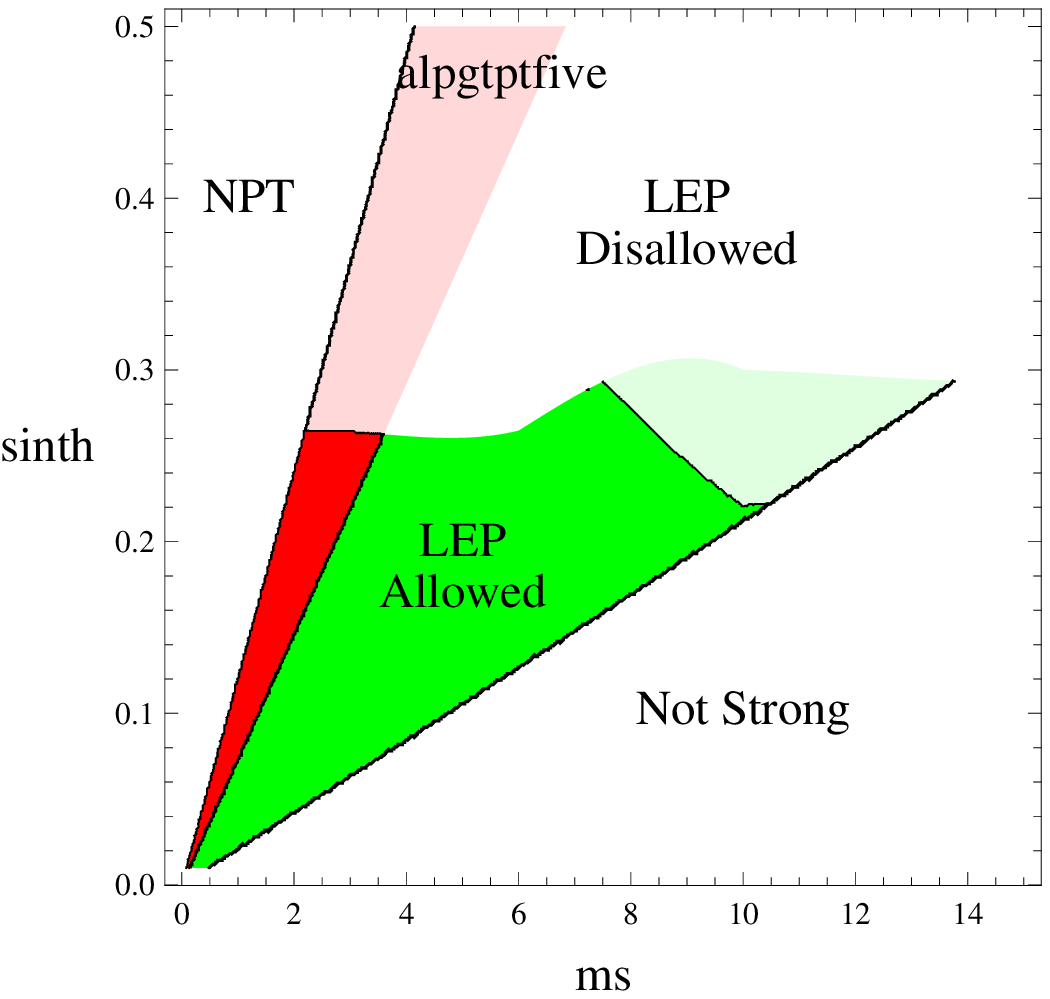}
   \hspace{0.5in}
   \includegraphics[width=2.5in]{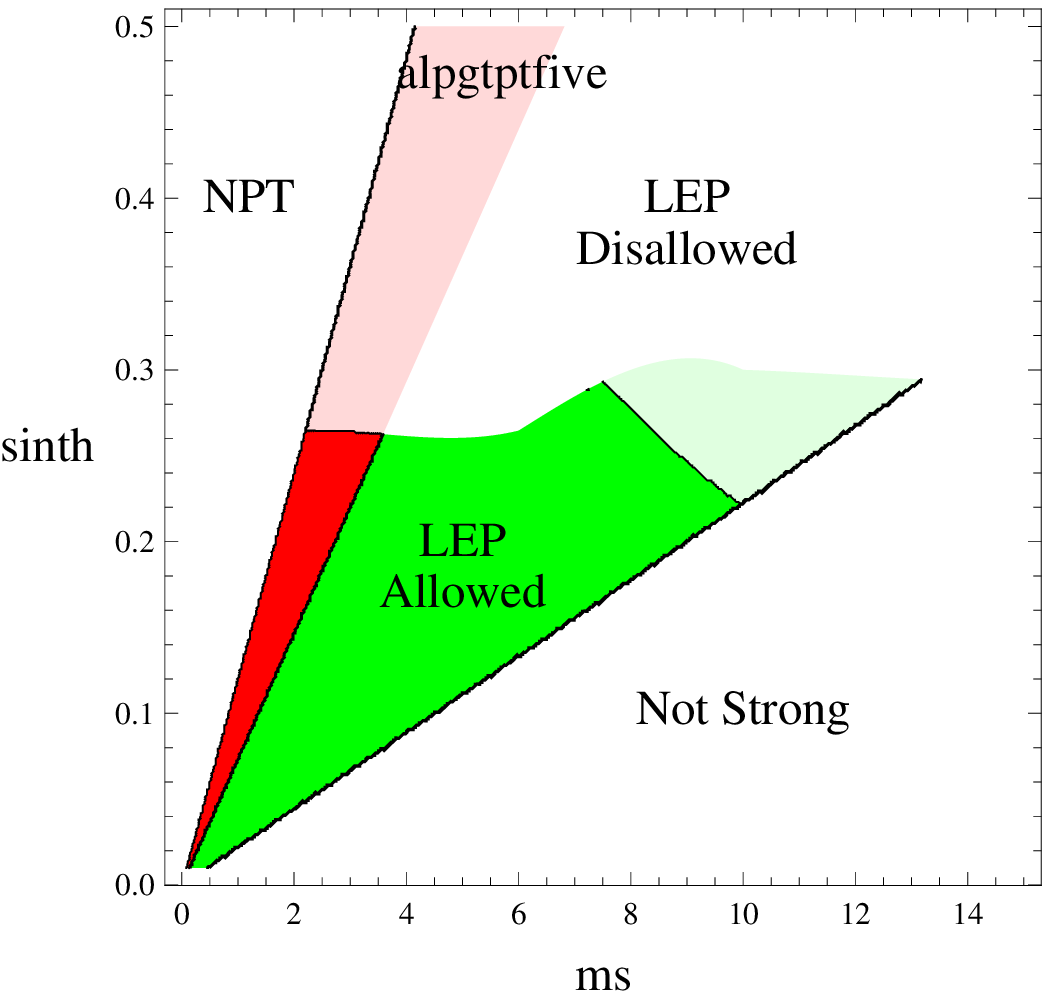}
 \caption{For $\overline{m}_H=90\,\gev$ (left) and $\overline{m}_H=115\,\gev$ (right) we plot the region allowed by LEP constraints and the requirement of a strong enough first order phase transition.  The dark green (larger) region assumes the singlet decays to jets or $\tau$'s and so applies the strongest LEP bound whereas the light green (smaller) region assumes the singlet decays some other way and applies only the model independent bound.  NPT denotes that for these parameters there is no phase transition, and the red (diagonal) band is the region where the latent heat released during the phase transition is sizeable.}
   \label{fig:allowedregions}
\end{figure}

The mixing between the singlet and the Higgs, due to $\kappa$, means that both mass eigenstates have couplings to $Z$ and so have strong constraints from LEP Higgsstrahlung processes~\cite{Barate:2003sz,Abbiendi:2002qp}.  The same coupling is also responsible for the decays of the mostly-singlet eigenstate.  Thus, for a singlet mass above approximately $12\,\gev$ the dominant decay will be to $b\bar{b}$ while below this threshold it will decay mainly to $c\bar{c}$ with a smaller branching ratio to $\tau^+\tau^-$.  There is the additional possibility that the singlet may have other couplings that do not affect the Higgs, and the discussion so far, but do alter the decays of $s$ in such a way that neither of the above bounds apply.  In this case the model independent \cite{Abbiendi:2002qp} Higgs bound applies; note this bound only applies for states lighter than $82\,\gev$.  Taking these bounds into account and requiring that (\ref{eqn:vevs}) leads to a sufficiently strong phase transition, satisfying (\ref{eqn:orderconstraint}), we find the allowed regions of parameter space.  These are shown in Figure~\ref{fig:allowedregions} for physical Higgs masses above and below the LEP bound.

The mostly-Higgs mass eigenstate has a new decay mode, into $2s$, which raises the intriguing possibility that the Higgs may have evaded the LEP bound \cite{Chang:2006bw}.  To allow for a Higgs much lighter than $114.4\,\gev$ the branching ratio to $b$ quarks must be greatly suppressed, for this reason we concentrate on a singlet lighter than 12 GeV. 
Such a scenario of a strongly first order phase transition with a light singlet
emerges in a model defined by (\ref{eqn:zerotemppot}).  While the potential in (\ref{eqn:zerotemppot}) does not yield an 80\% BR of $h\rightarrow ss$, which is needed to appreciably lower the LEP limit on the Higgs mass, a small additional $s^2 h^\dagger h$ operator would achieve this \cite{Chang:2006bw}.

In the simplest model we have considered, there is a considerable bound for $m_s\lsim 5$ GeV from $B\rightarrow X s$ decays, where $s\rightarrow \mu^+\mu^-$~\cite{Willey:1982mc,Grinstein:1988yu,Chivukula:1988gp,O'Connell:2006wi}. Indeed, the light region where $\alpha$ is large enough that a gravity wave signal is possible, would seem to be excluded. As we shall see momentarily, in general there is both a scalar ($s$) and pseudoscalar ($a$) added to the theory. While the scalar mixes with the Higgs, the pseudoscalar can easily remain light, offering a new decay channel, which can dominate over the Yukawa-suppressed decays without much difficulty. In this case, we have $s\rightarrow a a$, with $a\rightarrow \mu \mu $ or $a \rightarrow e e$, which are far more weakly constrained.

\section{Connecting to the Dark Sector}
\label{sec:susy}

In addition to having renormalizable couplings to the SM Higgs a singlet such as $s$ may have relevant couplings to new fields that are neutral under the SM.  If the new fields carry a conserved charge then they are stable and act as DM.  This is appealing since recent excesses in cosmic ray measurements \cite{Adriani:2008zr,Adriani:2008zq,Finkbeiner:2003im,Finkbeiner:2004us,Dobler:2007wv,Collaboration:2009zk} can be explained by a DM annihilating predominantly to leptonic final states.  The results suggest a DM mass of $\gtap\mathcal{O}(1\tev)$ and an annihilation cross section substantially larger than that expected from WIMP freeze out.  This enhancement of the annihilation cross section can be understood from a Sommerfeld enhancement from some new, light force carrier \cite{ArkaniHamed:2008qn,Pospelov:2008jd}\footnote{The Sommerfeld enhancement was first explored in the context of dark matter in \cite{Hisano:2004ds}, arising from weak interactions for multi-TeV WIMPs.}, while the hard lepton spectrum (without anti-protons) can be realized from the new mediator \cite{Cholis:2008vb,Cholis:2008qq}.  This mediator is typically taken to be a vector but here the light scalar fills this role.  Indeed, our model will be very similar in phenomenology to that of \cite{Nomura:2008ru}, in that the natural annihilation channel is $\chi \chi \rightarrow a s$.

However, a light scalar coupled to a massive field will not remain light under quantum corrections, unless there is a symmetry protecting its mass.  If the DM sector is highly supersymmetric then, as we will show, the scalar mass remains small.  In supersymmetry the coupling between the singlet and the Higgs may be generated from NMSSM type couplings such as $SH_u H_d$ in the superpotential.  Generically, since supersymmetry is broken in our sector, this will lead to a large mass for the scalar.  Instead we couple to the scalar using the ``supersoft'' operators $\int d^2 \theta\, W'_\alpha W^\alpha S$ and $\int d^2\, \theta W'_\alpha W'^\alpha S^2$ \cite{Dine:1992yw,Fox:2002bu} which generate masses for the scalars, as well as a Dirac mass between the singlino $\tilde s$ and the bino $\tilde B$, but do not generate large scalar masses through radiative effects once supersymmetry is broken.  Here $W'=\theta^2 D'$ is a spurion for SUSY breaking.  Such supersoft operators also generate the Higgs-singlet coupling necessary to raise the strength of the electroweak phase transition.  Including the DM mass term, the relevant part of the superpotential is,
\be\label{eq:relevantW}
W=M\,\chi\chi + \lambda S \chi \chi +  \frac{\kappa}{M} W'_\alpha W^\alpha S + \frac{\tilde{\kappa}}{M^2} W'_\alpha W'^\alpha S^2+ h.c.
\ee
The $\kappa$ term generates the singlet-Higgs coupling term of (\ref{eqn:zerotemppot}), as well as a Dirac mass between the singlino $\tilde s$ and the bino $\tilde B$.  Both $\kappa$ and $\tilde{\kappa}$ terms lead to scalar mass terms; $\kappa$ generates a mass for $\mathrm{Re}(s)$, naturally of the same size as the singlet-Higgs coupling, and $\tilde{\kappa}$ splits the scalar and psuedoscalar masses. The coupling of the DM to the singlet is responsible for the DM annihilation and the cross section is given by,
\be\label{eq:dmannihilation}
\langle \sigma_{ann} v\rangle \sim \left(\frac{|\lambda|}{0.5}\right)^4 \left(\frac{1000~\gev}{M_{\chi}}\right)^2 6\times10^{-26}\mathrm{cm}^3\mathrm{s}^{-1}~.
\ee
The correct relic abundance is achieved if $|\lambda|\sim 1/2$.   

We assume that in the MSSM sector there are additional sources of SUSY breaking which generate, amongst others, a Majorana mass term for the $U(1)$ gaugino, but that these sources of SUSY breaking are not coupled to the dark sector.  This arrangement, where the dark sector feels one source of supersymmetry breaking (supersoft breaking) and the MSSM experiences another (F-term breaking) can occur through sequestering.  If the MSSM is constrained to lie on the same brane as the source of supersymmetry breaking and the dark sector lives on a separate brane with the $U(1)'$ that acquires a D-term propagating in the bulk then the only source of supersymmetry breaking felt by the dark sector would come through supersoft operators~\cite{Carpenter:2005tz}.

In general the parameters in (\ref{eq:relevantW}) are complex, but by rephasing of the fields we may work in a basis where the Majorana gaugino mass term, the dark matter mass $M$, and $\kappa$ are real.  This has the advantage that it does not alter the discussion of previous sections.  The remaining phases, in $\lambda$ and $\tilde{\kappa}$, make the singlet scalar mass eigenstates an admixture of the scalar and psuedoscalar and the state that couples to the DM is no longer the same combination that couples to the Higgs.

The light scalars mix with the Higgs and thus have couplings to SM fermions which are essential to enable these states to decay before BBN.  If these scalars also have sizable coupling to the DM, as is needed for a large Sommerfeld enhancement to the annihilation cross section, then one would expect a large DM-nucleon effective coupling and an observable rate for DM-nuclear recoils.  For mixings of the size discussed earlier, $\sin\theta \sim \mathcal{O}(0.1)$, the DM-scalar coupling must be $\ltap 10^{-3}$.  However, the phases of the parameters in (\ref{eq:relevantW}) can be chosen such that this occurs and the decay rates, and annihilation rates, are large but the scattering cross section of DM off nuclei are small, thus evading current direct detection limits.  This can happen, for instance, if the CP symmetries in the SM and dark sectors are not aligned, the necessary smallness of the DM-scalar coupling is achieved if the two symmetries are almost anti-aligned.  The chiral superfield $S$ contains a complex scalar, $s+i a$, and with the couplings $\tilde{\kappa}$ and $\lambda$ such that one is almost purely real and one is almost purely imaginary, the CP-even eigenstate in the dark sector will be orthogonal to the CP-even eigenstate in the SM sector.  Thus, the state that has large mixing with the Higgs, $s$, will not have a large coupling to the DM, and instead $a$, which has small mixing with the Higgs, is responsible for the Sommerfeld enhancement.  It would be interesting to arrange for this to occur dynamically.

The singlino-bino system contains one light state, comparable to the singlet mass, which is mostly singlino with a small admixture $\theta \sim \kappa D'/(M M_{\tilde{B}})$ of bino.  In models with high scale supersymmetry breaking this would be the visible sector LSP and would result in too much DM and overclosure of the universe.  In models of low scale supersymmetry breaking with a light gravitino the singlino will decay, long before BBN, to a gravitino and a scalar singlet.  The scalar superpartners of the DM are also stable since they are charged under the DM-parity.  For the case of low scale breaking of supersymmetry they will also decay down to the DM through emission of a gravitino.

At freeze-out the annihilation cross section of the DM is given by (\ref{eq:dmannihilation}).  Under the assumption that dark matter is the fermionic component of $\chi$, the dominant annihilation is $\chi \chi \rightarrow s a$. For $m_s > 5 \gev$, we can have $s \rightarrow {\rm hadrons}$, while for lighter $s$, we assume $s \rightarrow a a$, in order to evade B-meson decay limits. The pseudoscalar will decay through its CP-violating mixing $a\rightarrow f \bar f$. At late-times, and low velocities, this annihilation may be enhanced by a DM-DM attractive potential induced by the exchange of the light scalar singlets, such a Sommerfeld enhancement requires $\overline{m}_S\ltap \lambda^2 M_\chi/4\pi$.   Amazingly these requirements coming from the dark sector are compatible with the requirement of a first order phase transition and are consistent with LEP constraints. Thus, the sector responsible for the FOPT can easily be the ``new force'' needed for Sommerfeld models, and in particular for models along the lines of \cite{Nomura:2008ru}.

\section{Bubble Nucleation}
\label{sec:bubble}

First-order phase transitions proceed via nucleation of bubbles of the broken phase within the symmetric phase.  The subsequent collisions of bubbles and the interactions of the bubble walls with the surrounding plasma are a source of gravitational radiation.  In this section, we study the possibility of gravity wave detection, by future experiments such as LISA~\cite{lisa} and BBO~\cite{bbo}, in our model.

The gravitational wave spectrum is characterized by two quantities $\alpha$ and $\beta$ \cite{Kamionkowski:1993fg,Apreda:2001us,Grojean:2006bp}.  $\alpha$ is the ratio of the vacuum energy density liberated at the phase transition, $\epsilon$, to the energy density of the symmetric phase, commonly radiation, thus $\alpha = \epsilon / \rho_{rad}$.   $\beta$ determines the size of bubbles at the time of collision and thus the characteristic frequency of the gravitational radiation, $f_{*}.$   It is determined from the rate of variation of the bubble nucleation rate, $\beta\approx d\log \Gamma /dt$.  The duration of the phase transition is given by $\beta^{-1}.$

In terms of these two quantites the energy density in gravity waves, from the phase transition is~\cite{Kamionkowski:1993fg},
\be
\label{eqn:omegagw}
\Omega_{GW}h^{2} \approx 1.1 \times 10^{-6} \kappa^2 \left({H_{*}\over \beta}\right)^{2} \left({\alpha\over 1+\alpha}\right)^{2} \left({v^{3}\over 0.24+v^{3}}\right) \left({100\over g_{*}}\right)^{1/3}~,
\ee
where $\kappa$ is the fraction of vacuum energy that is transferred to the bulk motions of the plasma and for the ranges of $\alpha$ we are interested in $\kappa\approx 1/2$.
The peak frequency of the waves occurs at,
\be
\label{eqn:freqgw}
f_{max} \approx 5.2 \times 10^{-8} Hz \left({\beta \over H_{*}}\right) \left(\frac{T_{*}}{1\, \gev}\right) \left({g_{*} \over 100}\right)^{1/6}~.
\ee
LISA has sensitivity ranging from $\Omega_{GW}h^{2} \gtap 2\times10^{-11}$ at $f\approx 3\times 10^{-4}$ Hz to $\Omega_{GW}h^{2} \gtap 2\times10^{-12}$ at $f\approx10^{-2}$ Hz while the sensitivity drops off considerably outside this range~\cite{lisa}.

Calculating both parameters requires determining the field profiles during the tunnelling process, over a range of temperatures; a difficult numerical problem for more than one field.  However, inspection of the potential in our case shows that there is a path between the two minima, along which $s$ is given by its vev $s=-\kappa(\phi^2+T^2/12)/m_s^2$, which minimizes the potential.  We assume that the tunneling process will follow this path in field space and this reduces the search for the bubble action, $S_3$, to a one field problem that can be solved by the usual methods~\cite{Coleman:1977py}.

\begin{figure}[ht] 
   \centering
   \psfrag{sinth}[][][1.]{$\sin\theta$}
   \psfrag{ms}[][][1.]{$\overline{m}_S$}
   \psfrag{0.0}[][][.8]{$0$}
   \psfrag{0.1}[][][.8]{$\!0.1$}
   \psfrag{0.2}[][][.8]{$\!0.2$}
   \psfrag{0.3}[][][.8]{$\!0.3$}
   \psfrag{0.4}[][][.8]{$\!0.4$}
   \psfrag{0.5}[][][.8]{$\!0.5$}
   \psfrag{0}[][][.8]{$0$}
   \psfrag{2}[][][.8]{$2$}
   \psfrag{4}[][][.8]{$4$}
   \psfrag{6}[][][.8]{$6$}
   \psfrag{8}[][][.8]{$8$}
   \psfrag{10}[][][.8]{$10$}
   \psfrag{12}[][][.8]{$12$}
   \psfrag{14}[][][.8]{$14$}
   \psfrag{fmax}[][][.8]{$f_{max}$}
   \includegraphics[width=2.5in]{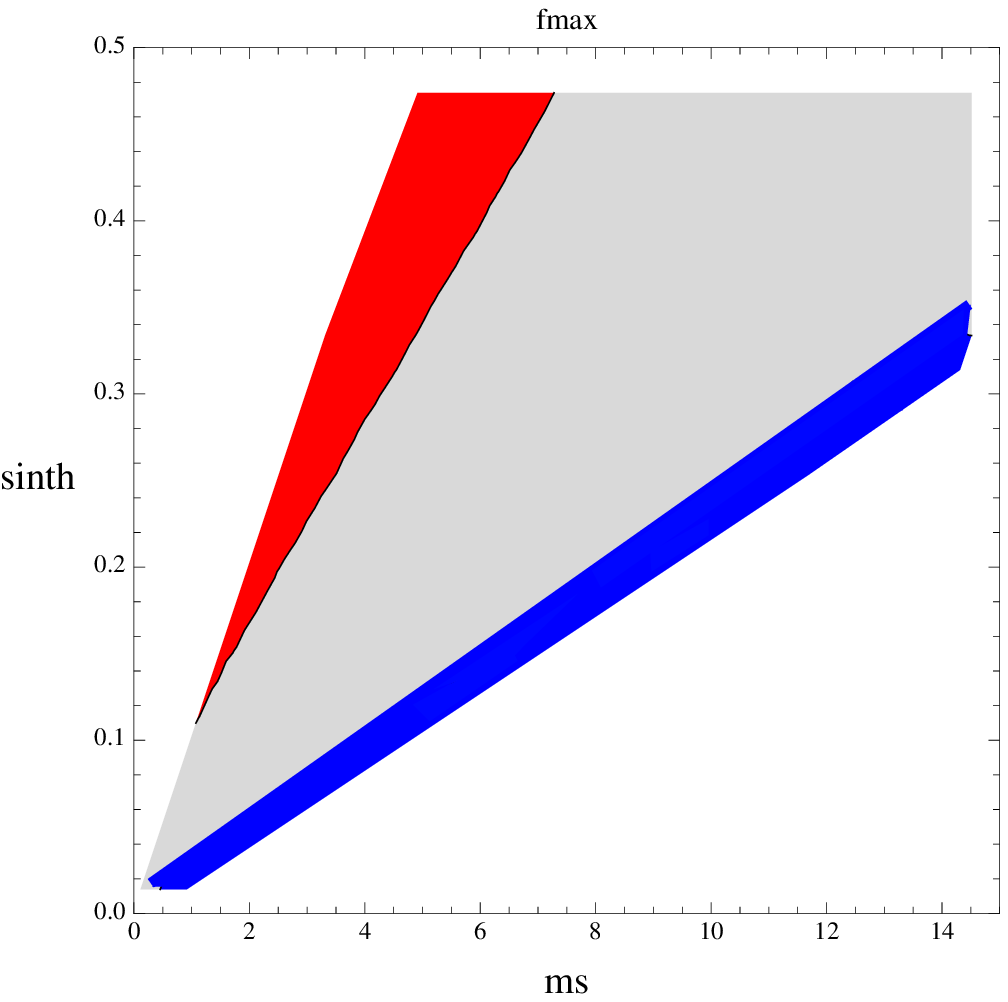}
   \hspace{0.5in}
   \psfrag{sinth}[][][1.]{$\sin\theta$}
   \psfrag{ms}[][][1.]{$\overline{m}_S$}
   \psfrag{0.0}[][][.8]{$0$}
   \psfrag{0.1}[][][.8]{$\!0.1$}
   \psfrag{0.2}[][][.8]{$\!0.2$}
   \psfrag{0.3}[][][.8]{$\!0.3$}
   \psfrag{0.4}[][][.8]{$\!0.4$}
   \psfrag{0.5}[][][.8]{$\!0.5$}
   \psfrag{0}[][][.8]{$0$}
   \psfrag{2}[][][.8]{$2$}
   \psfrag{4}[][][.8]{$4$}
   \psfrag{6}[][][.8]{$6$}
   \psfrag{8}[][][.8]{$8$}
   \psfrag{10}[][][.8]{$10$}
   \psfrag{12}[][][.8]{$12$}
   \psfrag{14}[][][.8]{$14$}
   \psfrag{omegahsq}[][][.8]{$\Omega_{GW}h^2$}
   \includegraphics[width=2.5in]{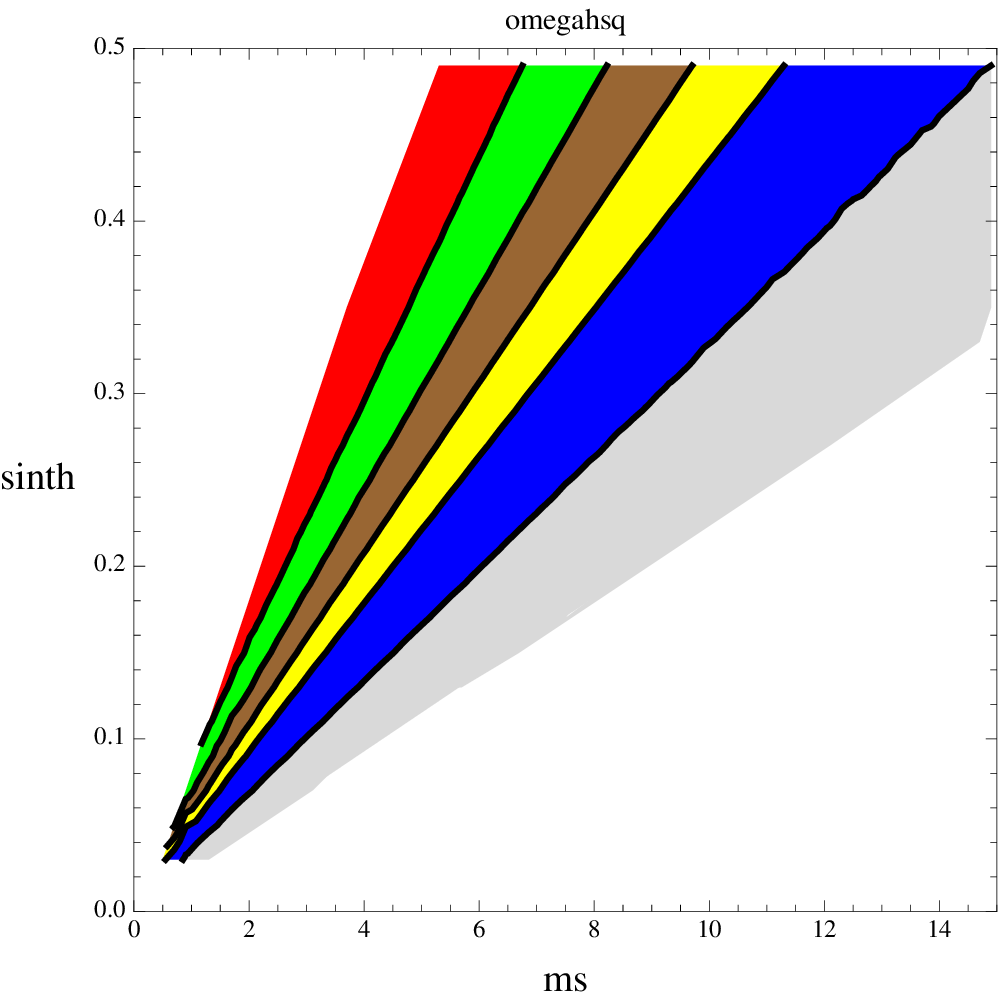}
   \caption{Plots of $f_{max}$ and $\Omega_{GW}h^2$ for $\overline{m}_H=115\,\gev$.  In the $f_{max}$ plot, the red (upper) region is in the sensitivity band of LISA and the blue (lower) in that of BBO.  In the $\Omega_{GW}h^2$ plot, reading from the top, the contours correspond to $10^{-15}$, $10^{-16}$, $10^{-17}$, $10^{-18}$, and $10^{-20}$ respectively.}
   \label{fig:GWplots}
\end{figure}

Finding the bubble profiles for $\phi$ and $s$ we are able to determine the temperature, $T_*$, at which the bubble nucleation rate is comparable to the Hubble expansion, i.e. $T_*^4e^{-S_3/T_*}\sim H^4$, once this is satisfied the bubbles are guaranteed to percolate.  At this time the bubble action is given approximately by $S_3\approx S_0-\beta(t-t_*)$, so once $t_*$ is reached the transition ends approximately time $\sim 1/\beta$ later, and the bubble size at this time is $R\sim \beta^{-1}$.  With these bubble profiles, and transition temperature in hand we can determine $\alpha$ and $\beta$ and thus $\Omega_{GW}h^2$ and $f_{max}$, which are shown in Figure~\ref{fig:GWplots}.  Although, in part of parameter space, the frequency is in the correct range to be seen by LISA the amplitude of the gravity waves is too small.  This is due, in part, to the fact that the phase transition happens late, as discussed earlier, consequently $T_*$ is low.  The transition happens close to $T_c$ where $\beta/H$ is large, pushing the frequency into the observable ranges, despite this low $T_*$, but at the expense of lowering the amplitude: the amplitude of the signal is determined by number of bubbles available for collision and the number of bubbles per Hubble volume is given by $\beta^{-1}\Gamma/H^3\sim H/\beta$.  Even for BBO the signal is not large enough in the frequency range to which they are sensitive; BBO's sensitivity range is estimated to be $\Omega h^2\gtap 2\times 10^{-17}$ over the range $0.1\,\mathrm{Hz} \ltap f \ltap 1\,\mathrm{Hz}$~\cite{Corbin:2005ny}.

\section{Conclusions}
\label{sec:conclusions}
New data from cosmic rays may suggest the presence of a new scale of physics with mass $\sim$ GeV. Such states may also allow a lighter Higgs boson consistent with LEP limits by modifying its decays. If the new state $s$ couples to the Higgs with a trilinear coupling, it can naturally generate a first-order electroweak phase transition, which is a necessary ingredient for electroweak baryogenesis.

Such a state naturally mixes with the Higgs boson. LEP limits on $s$-strahlung then naturally constrain its properties. We have found that the constraints are easily satisfied if the state's mass is $m_s \lsim 12$ GeV, where its decays into bottom quarks are small. Allowing fully general decays of $s$ opens up regions of parameter space below 5 GeV.
 While such a scalar can generate a first-order phase transition, the duration of the transition is sufficiently small that no appreciable gravitational wave signal arises.

New searches~\cite{Essig:2009nc,Bjorken:2009mm} for light states will add insight into whether this scenario is realized in nature. However, in light of the many motivations for new physics at the $\sim$ GeV scale, it is intriguing that such particles may also open the window for the wide ranging phenomenology associated with a first order phase transition.

\section*{Acknowledgements} We thank Zoltan Ligeti, Gilad Perez, and Mark Wise for discussions. NW and AK are supported by NSF CAREER grant PHY-0449818 and DOE OJI grant \# DE-FG02-06ER41417.  Fermilab is operated by Fermi
Research Alliance, LLC, under Contract DE-AC02-07CH11359 with the United States
Department of Energy. The research of SD is supported by the Natural Sciences and Engineering Research Council of Canada.  PF and NW would like to thank the University of Washington's Workshop on Signatures of Long-Lived Exotic Particles at the LHC where part of this work was completed; this work was supported in part by the DOE under Task TeV of contract DE-FGO2-96-ER40956.
\bibliography{foptbib}
\bibliographystyle{apsrev}

\end{document}